\documentclass[review]{elsarticle}

\usepackage{fullpage}
\usepackage{blindtext}
\usepackage{amsmath}
\usepackage{mathrsfs}
\usepackage{amssymb}
\usepackage{url}
\usepackage{color,xcolor}
\usepackage{braket}
\usepackage{appendix}
\usepackage{multirow}
\usepackage{adjustbox}
\usepackage{graphicx}
\usepackage{array}
\usepackage{longtable}
\usepackage{natbib}
\usepackage{amsfonts}
\usepackage{physics}
\usepackage{subfigure}
\usepackage{acronym}
\usepackage{diagbox}
\usepackage[amsmath,standard,thmmarks]{ntheorem} 
\usepackage{microtype}

\acrodef{ML}[ML]{Machine Learning}
\acrodef{SCD}[SCD]{Sickle cell disease}
\acrodef{RBCs}[RBCs]{Red Blood Cells}
\acrodef{KNN}[KNN]{K-nearest Neighbour}
\acrodef{ANN}[ANN]{Artificial Neural Network}
\acrodef{CNN}[CNN]{Convolutional Neural Network}

\journal{}

\begin{document}

\begin{frontmatter}

\title{A Novel Deep Learning based Model for Erythrocytes Classification and Quantification in Sickle Cell Disease}


\author[UD]{Manish Bhatia}
\ead{manishbhatia055@ducic.ac.in}

\author[UD]{Balram Meena}
\ead{Balram4327@gmail.com}

\author[vipin]{Vipin Kumar Rathi\corref{mycorrespondingauthor}}
\ead{vipin68\_scs@jnu.ac.in}

\author[halmstad,aalto]{Prayag Tiwari\corref{mycorrespondingauthor}}
\ead{prayag.tiwari@ieee.org}

\cortext[mycorrespondingauthor]{Corresponding author}

\author[leeds]{Amit Kumar Jaiswal}
\ead{amitkumarj441@gmail.com,a.jaiswal@ucl.ac.uk}

\author[csir]{Shagaf M Ansari}
\ead{shaghaf@iiim.res.in}

\author[csir]{Ajay Kumar}
\ead{ajaykumar@iiim.res.in}

\author[aalto]{Pekka Marttinen}
\ead{pekka.marttinen@aalto.fi}

\address[UD]{Cluster Innovation Centre, University of Delhi, New Delhi, 110007, India.}
\address[vipin]{Department of Computer Science, Ramanujan College,  University of Delhi, New Delhi, 110019, India.}
\address[leeds]{University College London, United Kingdom}
\address[halmstad]{School of Information Technology, Halmstad University, Sweden}
\address[aalto]{Department of Computer Science, Aalto University, Finland}
\address[csir]{CSIR-Indian Institute of Integrative Medicine, Canal Road, Jammu-180001, India}

\begin{abstract}
The shape of erythrocytes or red blood cells is altered in several pathological conditions. Therefore, identifying and quantifying different erythrocyte shapes can help diagnose various diseases and assist in designing a treatment strategy. \ac{ML} can be efficiently used to identify and quantify distorted erythrocyte morphologies. In this paper, we proposed a customized deep convolutional neural network (CNN) model to classify and quantify the distorted and normal morphology of erythrocytes from the images taken from the blood samples of patients suffering from \ac{SCD}.  We chose \ac{SCD} as a model disease condition due to the presence of diverse erythrocyte morphologies in the blood samples of SCD patients. For the analysis, we used 428 raw microscopic images of \ac{SCD} blood samples and generated the dataset consisting of 10, 377 single-cell images. We focused on three well-defined erythrocyte shapes, including discocytes, oval, and sickle. 
We used 18 layered deep CNN architecture to identify and quantify these shapes with 81\% accuracy, outperforming other models. We also used SHAP and LIME for further interpretability. The proposed model can be helpful for the quick and accurate analysis of \ac{SCD} blood samples by the clinicians and help them make the right decision for better management of \ac{SCD}. 

%what we found

\end{abstract}

\begin{keyword}
Erythrocytes, Intelligent System, Sickle cell disease, CNN, Interpretability, Classification

\end{keyword}

\end{frontmatter}

% \linenumbers

\section{Introduction}
\label{intro}
Investigation of erythrocyte morphology is an essential task for hematological diagnosis. Erythrocytes or red blood cells display different types of shapes and sizes under various pathological conditions such as sickle cell disease, thalassemia, megaloblastic anemia, iron deficiency, liver disorders, oxidative hemolysis, myelodysplasia, lead poisoning, etc. \cite{petrovic2020sickle}. Therefore, it becomes critical for a hematologist to identify the correct morphology to link it with a specific disease. Erythrocytes are filled with hemoglobin, which works to carry oxygen from lungs to tissue and carbon dioxide in the opposite direction. Any variation in the structure or content of hemoglobin can therefore alter the morphology of erythrocytes. Hemoglobin is a quaternary protein structure containing two chains, which are alpha-globin and beta-globin. All the polypeptide chains have heme at their ends. The ferrous atom present in the heme group can reversibly bind with oxygen and converting it into oxyhemoglobin. The binding of oxygen with ferrous atoms reduces the gap between two polypeptide chains, thus leading to the contraction of hemoglobin molecules. However, when oxygen is released from the hemoglobin molecule, it expands again. This remarkable ability of hemoglobin to bind and release oxygen makes it a unique molecule, contributing a lot to our understanding of biology and chemistry at a molecular level. The elucidation of hemoglobin by Max Perutz led him to the chemistry Nobel Prize in 1962 \cite{perutz1960structure}.

More than 200 variants of hemoglobin have been reported during the last few decades, though all the variants are not linked with a disease condition \cite{marengo2006structure}. However, many mutations in the alpha or beta chains of hemoglobin can lead to severe disease conditions like sickle cell disease and thalassemia. Both these diseases fall under the category of hemoglobinopathies, which involve the genetically inheritable abnormal structure of hemoglobin. Geographically, \ac{SCD} is very common throughout the African region, especially in the Sub-Saharan Africa. There is very high prevalence of \ac{SCD} among people of Africa. Strangely, SCD was first mentioned by Dr. Robert Herrick from the West-Indies in 1910 as 'sickle cell anemia' \cite{herrick1910peculiar}. Linus Pauling was the first to report the abnormality in the structure of hemoglobin of sickle hemoglobin (HbS) in 1949. Since then, a lot of research has been done on this disease condition. The adult hemoglobin consists of two alpha and two beta chains ($\alpha_2\beta_2$). In sickle cell disease, both the genes for beta chains carry a point mutation in the sixth codon, where adenine substitution by thymine leads to the replacement of glutamic acid to valine in the beta chain of the polypeptide. Valine at the sixth position under low oxygen conditions interacts with the hydrophobic pocket of the adjacent beta-globin polypeptide to form a polymer structure \cite{bunn1997pathogenesis}. Under severe oxygen deficiency conditions, the level of polymerization may further be increased to alter the morphology of erythrocytes to different shapes, including crescent or sickle shapes. These distorted erythrocytes have a fragile membrane and thus are very prone to hemolysis. The release of heme due to lysis may further activate the NLRP3 inflammasome to produce IL-1$\beta$ mediated inflammation. Furthermore, sickled erythrocytes cannot easily pass through small capillaries, leading to clogging and vaso-occlusion crisis. The clogging of blood vessels can lead to stroke or cause severe damage to different body organs, including the liver, spleen, kidney, etc. The goal of long-term clinical management of the disease is to keep the level of HbS $<30\%$ to 40\% to avoid vaso-occlusion crisis-like complications and to maintain hemoglobin level at 10 g/dl which helps conserve the oxygen-carrying capacity \cite{howard2016sickle}. However, under acute conditions, the goals of disease management may differ according to the presentation.
During the last few years, drugs such as L-glutamine, GBT440, and Crizanlizumab have been approved by US-FDA to treat \ac{SCD}. However, the treatment of \ac{SCD} still relies mainly on simple or exchange blood transfusion and hydroxyurea therapy. Both these treatment approaches are meant to keep the level of sickled erythrocytes low, along with maintaining the hemoglobin level at 10 g/dl. First, however, it is essential to decide when to give a blood transfusion. Under clinical setup, this is usually decided by the severity of the presentation of the disease or crisis. Recently, efforts have been made to identify different shapes of distorted erythrocytes in the blood of \ac{SCD} patients by using a deep neural approach \cite{xu2017deep}. In this study, we have developed a model to identify and quantify the three most common types of erythrocyte shapes from digital microscopy images of blood smears from \ac{SCD} patients. 

\subsection{Contributions}
The main contributions of this paper is as follows:

\begin{itemize}
\item A new \ac{SCD} dataset is created under the supervision of 50 medical staff at Government Medical College (GMC), Jammu, India.  
\item We proposed a custom DeepCNN model and compared it with several state-of-the-art models (LightGBM classifier, InceptionResNetV2, VGG16, VGG19, ResNet50, ResNet50V2, Fine-Tuned ResNet50V2, DenseNet121, and EfficientNet B0) for this study. 
\item The experimental results on the proposed model show a promising improvement compared to the other models for identifying and quantifying erythrocytes in \ac{SCD}.
\item Furthermore, LIME and SHAP is used for the proposed model interpretability.
% \item We believe that by using this architecture, full-fledged software can be developed. Using this method, the clinicians can easily quantify the distorted erythrocytes and can thus take the decision for appropriate treatment.
\end{itemize}

\subsection{Organization}
The following paper is organized as follows: In section \ref{sec:1}, we discuss related work and emphasize various approaches used. Then, section \ref{dataset} briefly describes the methods used for data collection. Subsequently, section \ref{sec4} explains the methodology used for the problem formulation and gives brief details about the algorithms, segmentation, object detection methods, along with the training of machine learning and deep learning models. Experimental results can be found in Section \ref{sec5}. Finally, discussion can be found in Section \ref{discussion} followed by the last section, which concludes the paper.

\section{Related Work}
\label{sec:1}
Manually classifying erythrocytes or \ac{RBCs} can be a tedious task even for medical experts/practitioners. To classify a blood sample manually, one has to inspect the blood sample using a microscope to examine each frame, which can be subjective and time-consuming.

These manual tasks can be automated effectively, especially using computer vision and deep learning methods. Deep learning (DL) methods are the most suitable methods for high-dimensional image datasets. These methods use high computational power, increasing the optimization methods and reducing the loss/error propagation while addressing other practical problems. Recently, there have been many improvements in medical imaging using DL approaches leading to very effective models.

A software package (RSIP Vision) \cite{ben2017retinal} is also available to recognize red blood cells using a \ac{ML} classifier to classify the morphological features. However, the procedure of data-processing can be very time-consuming as most of the annotations requires high-pixel images or a frame-level video annotation. In \cite{liu2014sparse}, the authors proposed a HEp-2 cell classification model using a superpixels-based sparse coding approach via transfer learning. Their method could stably transform the visual feature from low-level to high-level, indicating the samples and bases. The highest overall accuracy was reported as 95.4\%.

As discussed below, some researchers proposed algorithms based on deep learning to detect and classify sickle \ac{RBCs} based on blood sample images.
A study \cite{gonzalez2014red} on the detection of sickle cell disease was carried out where the level set method \cite{lin2004medical} was used. This method collects the image-region and image-boundary statistical information. It was developed to generate the outline of an image along with the concave point detection technique, which is used to detect the region of interest. Their proposed model achieved more than 98\% accuracy with 100\% efficiency while detecting an elongated red-blood-cell through the real image. But there were also some setbacks. The used input images in the whole process were not pre-processed, meaning the presence of WBCs (white blood cells) and platelets in images can cause false detection of sickle RBCs. \ac{KNN} classifier approach was used to classify \ac{RBCs} into sickle cells, dacrocytes, and ovalocytes \cite{sharma2016detection}. Their proposed model classifies the \ac{RBCs} with 80.6\% accuracy but fails to segment WBCs (white blood cells) from RBCs. In \cite{tomari2014computer}, an \ac{ANN}-based classifier is used for RBC classification in blood smear images. Their proposed model could classify non-overlap RBCs using ANN with 83\% accuracy but cannot classify a cluster of or overlapping RBCs. A novel approach for the separation of partially overlapping cells was developed based on shape information \cite{song2009new}. The result showed 87\% of the cluster as correct separation, 7\% of the clusters as under-separated, and 6\% as over-segmented clusters. A deep CNN was used for the classification of SCD, stating to have the capability of detecting five different distorted cells: Echinocytes, Discocytes or Oval, Elongated or Sickle, Reticulocytes and Granular \cite{xu2017deep}. The experimental results showed good performance robustly and high accuracy for all classes of distorted cells. In \cite{zhang2018rbc}, a U-Net framework was used to automate the semantic segmentation for SCD red blood cells. The baseline U-Net was fine-tuned by replacing the convolution kernel with deformable convolution. The results showed higher single-RBC segmentation accuracy.

\section{Method for Dataset Collection}
\label{dataset}
\subsection{Blood samples}
Human blood samples are collected from 10 patients diagnosed with Sickle Cell Anemia. The sample collection was performed under the supervision of medical staff at Govt. Medical College (GMC), Jammu. Prior informed consent was given by the volunteers before the collection of blood samples.  The Institutional Ethics Committee GMC, Jammu, approved all the study protocols with approval number IEC/GMC/Cat A/2020/267.

\subsection{Induction of Sickling}
The presence of sickling in erythrocytes samples obtained from SCA patients contained 7-8\% sickled RBCs. We enhanced the sickling level by treating them with sodium metabisulfite (SMB) for this study. Blood samples were incubated in HANKS’ balanced salt solution (HBSS) for 15 minutes in the presence of 1\% SMB in the carbon dioxide incubator. The images were taken at 40X on the phase-contrast microscope.

\subsection{Dataset}
\begin{figure*}
\centering
  \includegraphics[scale=0.40]{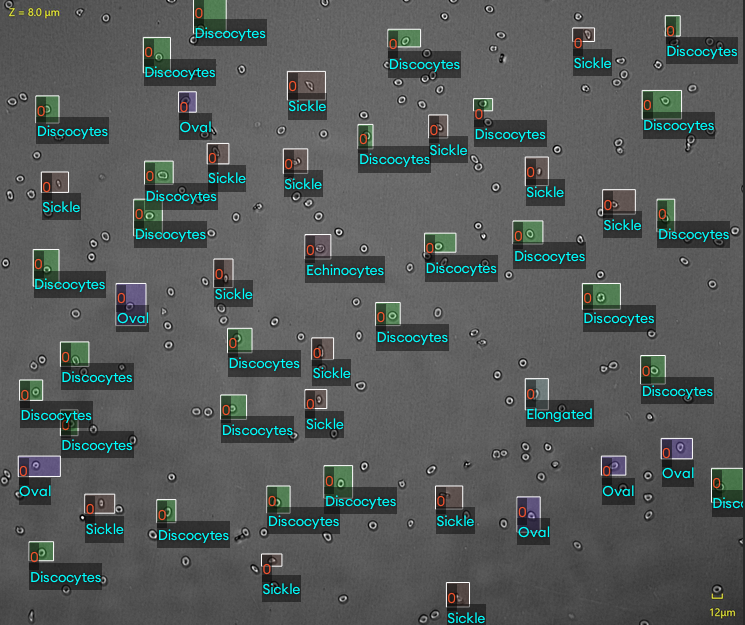}
\caption{Annotated image sample of Sickle \ac{RBCs}}
\label{fig2}
\end{figure*}

A total of 428 microscopic images of sickle \ac{RBCs} were extracted from 10 SCD patients. The width x height of the original image samples is 2528 x 2136. The images extracted were then annotated using SuperAnnotate software for the basic understanding of the cell types. A visualization of sickle \ac{RBCs} is shown in Fig \ref{fig2}. These image samples were then stored together in a way to make a reliable dataset for future use. 

\section{Proposed Methodology}
\label{sec4}

\begin{figure*}
\centering
  \includegraphics[scale=0.1]{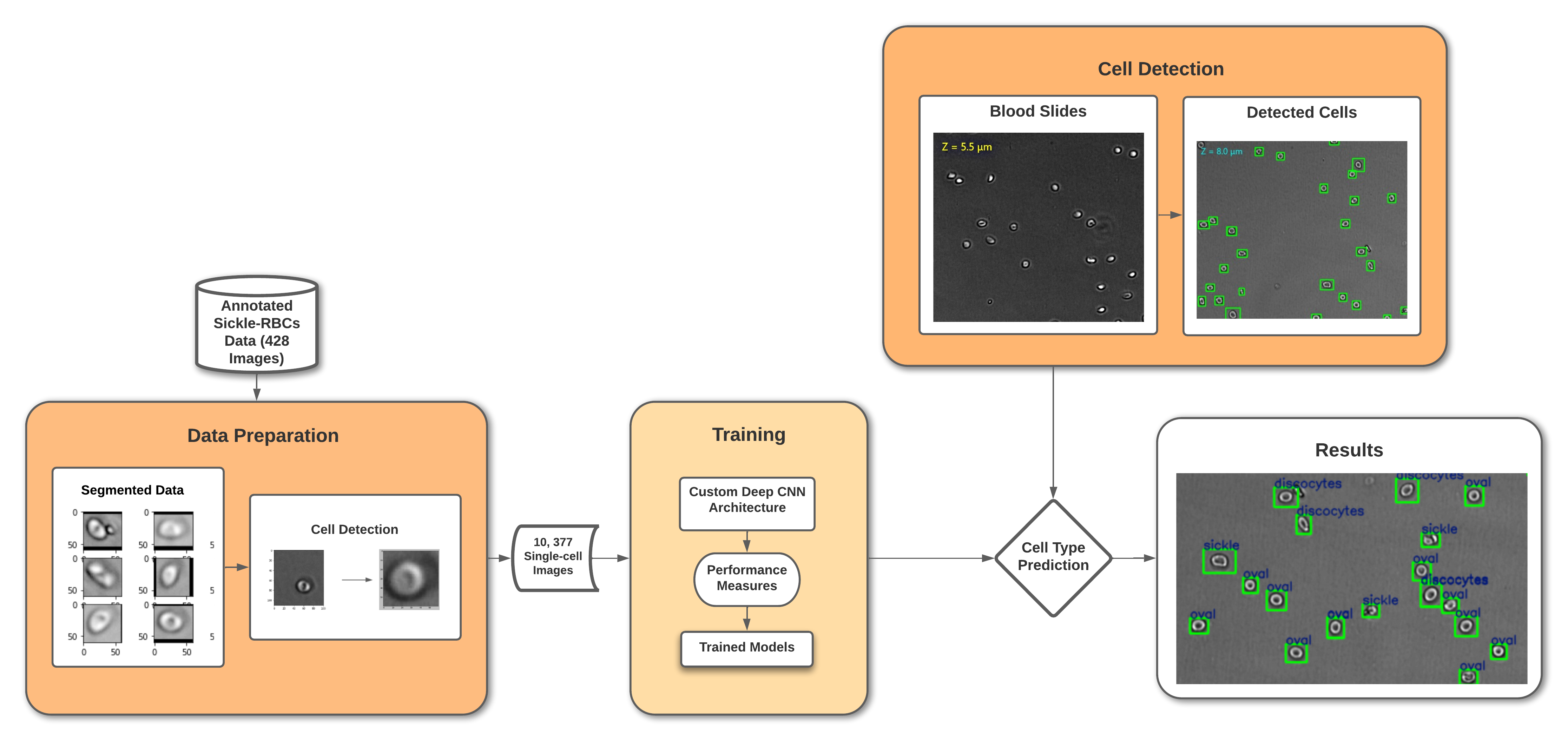}
% figures can also be resized by adding to the include graphics command as follows
% \includegraphics[width=0.5\textwidth]{images/proposed-framework-transparent.png}
\caption{The Proposed Framework}
\label{fig1}

\end{figure*}

\subsection{Problem Formulation}
Sickle cell anemia is an inherited red blood cells disorder where the RBCs cause vaso-occlusion phenomena for the human body due to distortion and inflexibility of cells. As SCD patients are always at risk of life-threatening complications, implementing cell detection methods and classifying them into their respective categories can enable medical technology to automate clinical diagnosis. 
In our work, we consider the task of identifying the ‘Sickle \ac{RBCs}’ through multi-class classification. However, there are several challenges: 1) We have a small unbalanced dataset of 428 images containing only 3 types of sickle RBCs. 2) In some images, RBCs overlap or appear as a cluster making it highly difficult to recognize every single cell. 3) Finally, the presence of artifacts and image shading in the images are also available.

\subsection{Algorithm}
Deep neural networks (DNNs) have been proved useful in many computer vision tasks. We provide details of the state-of-the-art techniques and a custom deep CNN model, alongside cell segmentation and cell detection. 

\begin{enumerate}
  \item LightGBM is a new boosting framework that uses tree-based learning, first introduced in 2017 \cite{ke2017lightgbm}. It works differently, faster, and better than most boosting methods. The same reason why it works is similar to state-of-the-art algorithms. When the dataset was fed to this framework, it showed promising results. We found that having too small a dataset, the model was not overfitting but instead lacked the ability to predict one of the three sickle RBCs classes.  
  
  \item We considered some pre-trained CNN architectures  such as InceptionResNetV2 \cite{kassania2021automatic}, ResNet \cite{kassania2021automatic} variants, DenseNet \cite{kassania2021automatic} and EfficientNet \cite{tan2019efficientnet} variants. All of these models work and perform differently from each other. However, we found that ResNets performed better in our classification problem. Residual networks had better training loss and accuracy than DenseNet and EfficientNet networks. Models such as VGG16 \cite{simonyan2014very}, and VGG19 \cite{simonyan2014very} couldn’t perform well, which is due to having a small dataset. From a performance perspective, out of all variants of ResNets, ResNet50V2 performed the best. Even though ResNet50V2 stands out to be significantly the best model in our case, the results weren't that promising, and the overall achieved accuracy was 50\%. We examined and found that the pre-trained models lacked the ability to recognize the images where RBCs were either overlapping or making clusters. We use the segmentation method to divide the overlapping cells or clusters of cells to solve this issue. Furthermore, we used object detection so that the model could identify these RBCs more significantly and precisely. Once the images were divided and cells were detected and stored into a single-cell image, we fine-tuned our ResNet50V2 model, which significantly improved the performance and results of the overall classification task as shown in Table \ref{table1}. 
  
  \item After computing the pre-trained models, a custom deep CNN-based model was proposed with 18 layers. After analyzing and comparing our custom model with Fine-tuned ResNet50V2, we found that the custom model had much better computational speed and performance metrics than the Fine-tuned ResNet50V2 model.
\end{enumerate}

\subsection{Cell Segmentation}\label{subsec:segmentation}
\begin{figure}
\centering
  \includegraphics[scale=0.6]{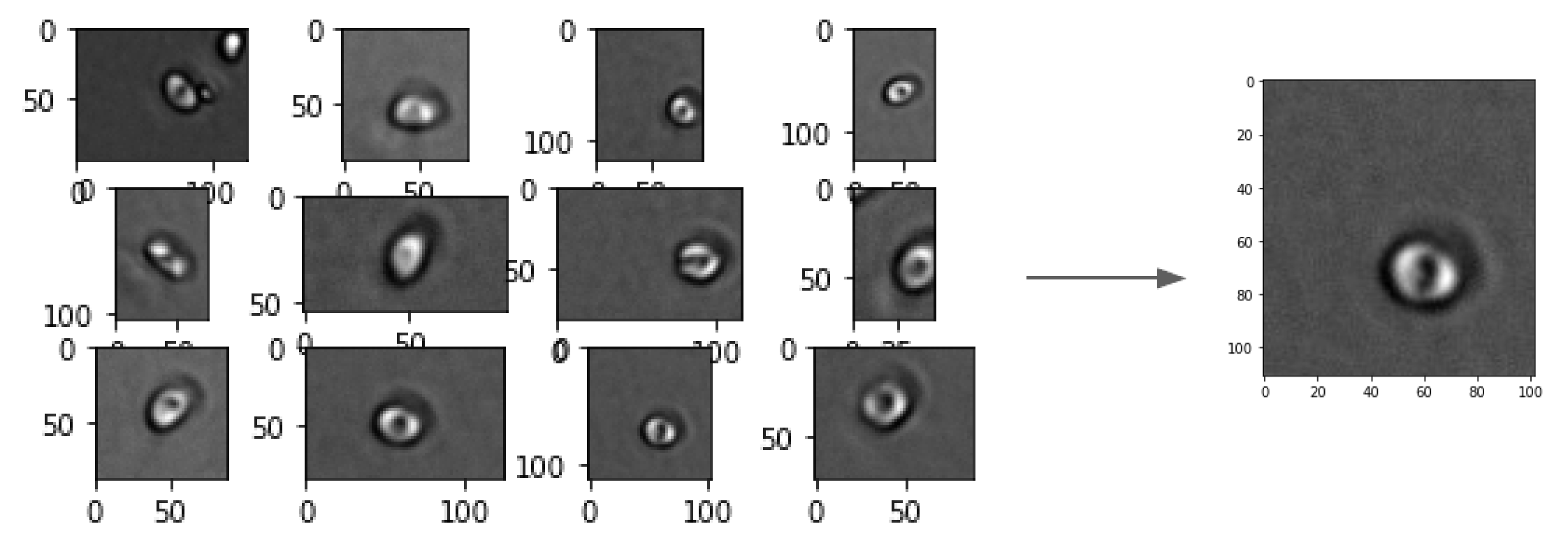}
\caption{Segments of a single image}
\label{fig3}   

\end{figure}

As mentioned earlier, the annotated image samples collected from the SuperAnnotate platform contain extra noise and overlapping cells that can be reduced using the cell segmentation approach. This approach is used to create several partitions or segments of microscopic images to a single-cell individual image. This segmentation approach or task is the foundation of image-based cellular research.
In our dataset, most of the Sickle \ac{RBCs} marked in a bounding box contain some extra space, due to which some cells are hard to process even with bounding boxes. To resolve this, we first divide our image samples into various small images that can also be called segments. Processing an entire image is not always a great idea as there are always some regions in the image that do not contain any information. To avoid this, the current dataset was divided into image samples, each containing exactly one known cell. All the annotated bounding boxes were split into several single-cell images.

\subsection{Cell Detection} \label{subsec:detection}
\begin{figure}
\centering
  \includegraphics[scale=0.6]{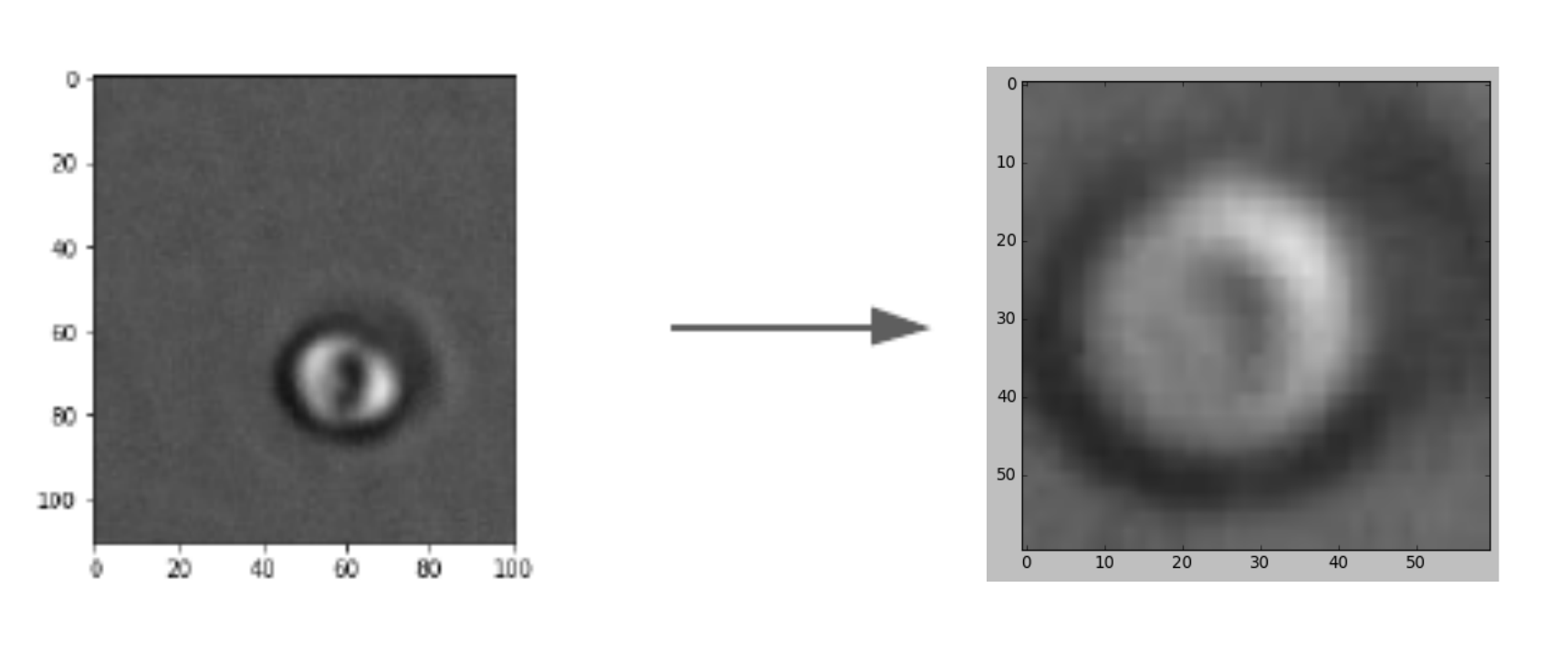}
\caption{Detected cell without any extra noise.}
\label{fig4}

\end{figure}

We now use the bounding box technique. Bounding box techniques are usually based on manual readings and are labor-intensive work. In the original 428 microscopic images, bounding boxes were created manually using SuperAnnotate software. Therefore, the generated images covered a large number of unwanted artifacts and noise, when fed for the training to CNN, reduced the accuracy and efficiency of the method. A visualization of an annotated bounding box with extra noise and artifacts can be seen in Fig. \ref{fig3}, which is obtained from SuperAnnotate. In addition, we cropped out all the bounding boxes and created several single-cell images to make further cell detection tasks more precise. 

\begin{figure}
\centering
  \includegraphics[scale=0.40]{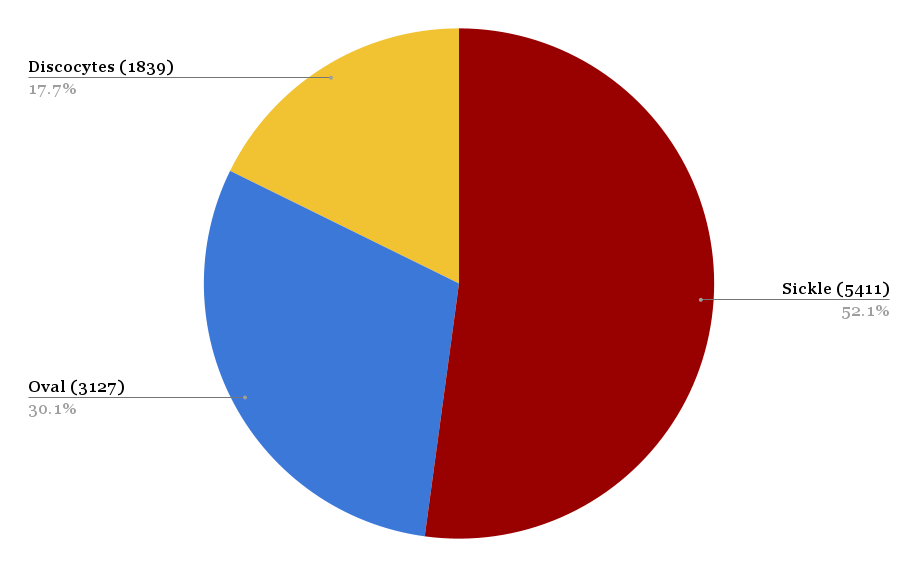}
\caption{Number of Images}
\label{fig5}
\end{figure}

Using the segmented images, we use the bounding box technique by applying the object detection technique for image transformation in OpenCV. Through this approach, we remove all the extra space, noise, and artifacts in the image to only detect the cell in the box (as shown in Fig. \ref{fig4}). Similarly, we do the same for the rest of the segmented images. After computing all the images, we create a new normalized dataset of single-cell images which can be used to train machine learning classifiers and deep CNN. The new dataset generated contains 10,377 single-cell image samples. These new single-cell image dataset were assigned to 3 different categories namely; sickle, discocytes, and oval. A visualization of the number of images for each type of cell can be seen in the Fig. \ref{fig5}. 

\subsection{Training}
In this section, we explain the detailed analysis of the training period of our proposed model along with the other different approaches: a) Light GBM, b) Pre-trained Deep Neural Network Models, and c) Proposed Custom Deep CNN Architecture.

\subsubsection{Light GBM (Light Gradient Boosting Machine)}
\begin{figure}
\centering
  \includegraphics[scale=0.3]{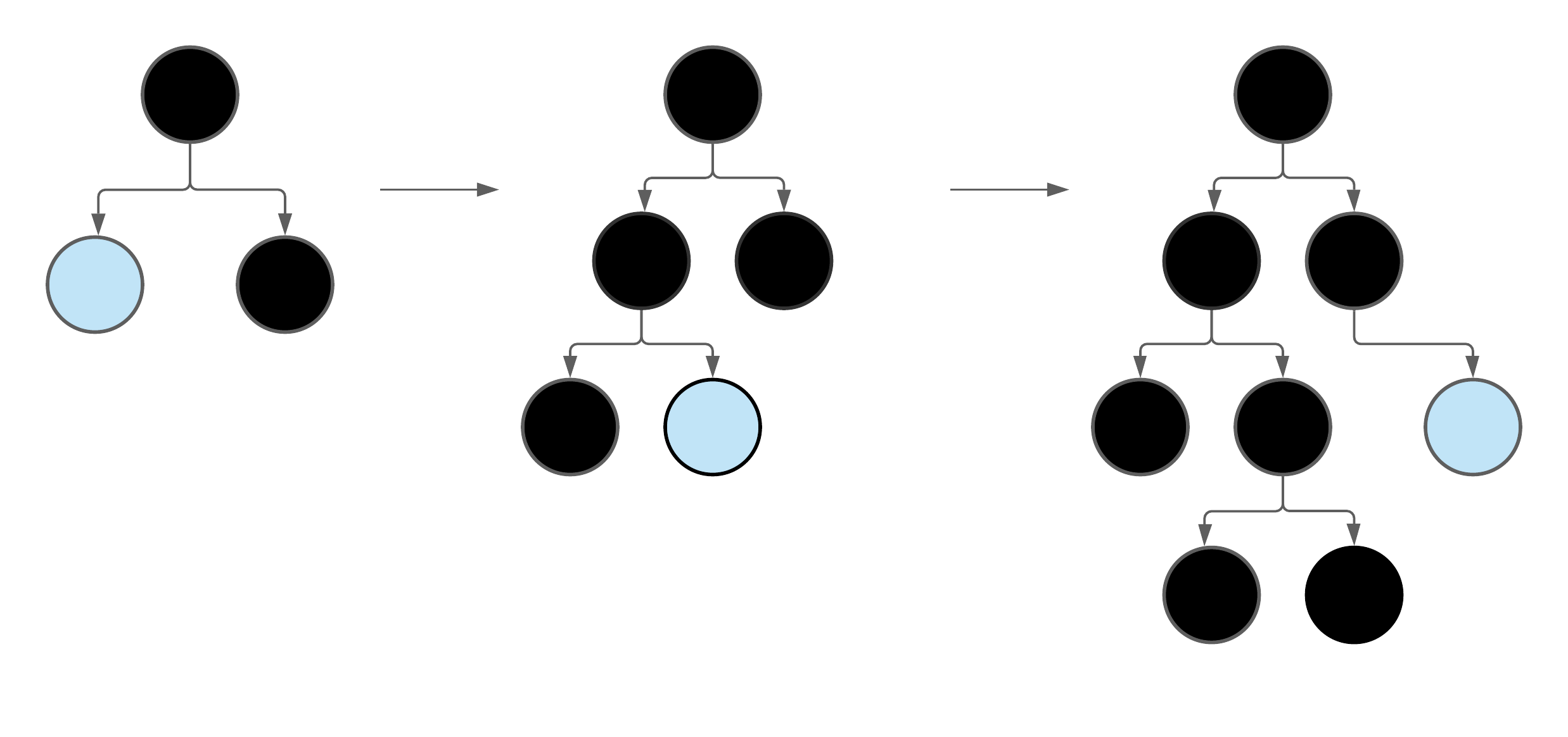}
\caption{LightGBM Structure (Vertically Growing)}
\label{fig6}

\end{figure}

\begin{figure}
\centering
  \includegraphics[scale=0.3]{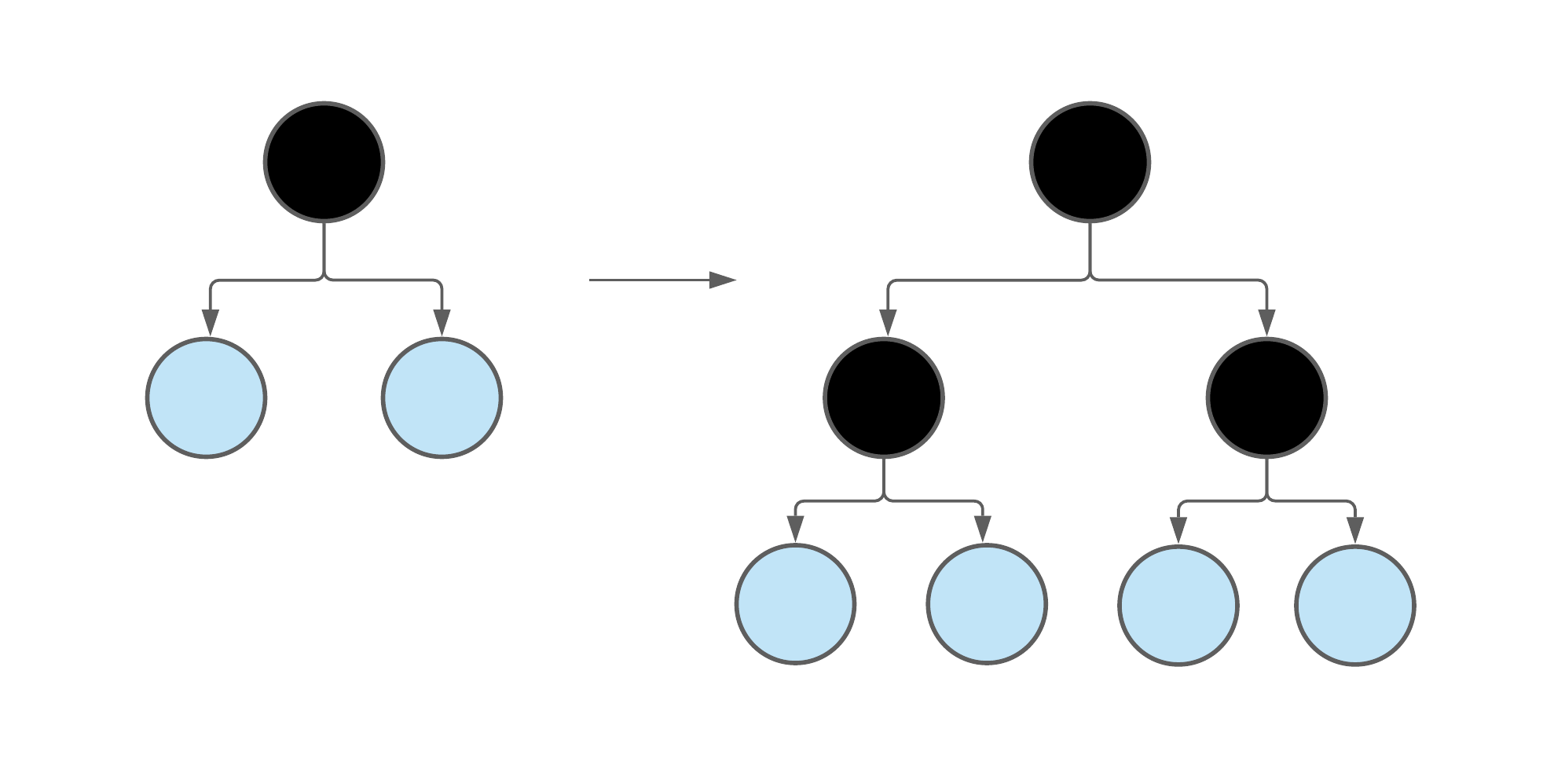}
\caption{Other Boosting Methods Structure (Horizontally Growing)}
\label{fig7}

\end{figure}

An algorithm developed by the researchers at Microsoft in 2017 proved to outperform other well-known boosting algorithms such as XGBoost and SGB (Stochastic Gradient Boosting) for memory consumption and impressive computational speed on big datasets \cite{ke2017lightgbm}. LightGBM is better than most boosting algorithms because it grows tree leaf-wise (vertically) while other boosting algorithms grow level-wise (horizontally). A visualization of how the working of LightGBM is different from others can be seen in Fig. \ref{fig6} and Fig. \ref{fig7}. 

% \begin{figure}
%      \centering
%      \begin{subfigure}[b]{0.20\textwidth}
%          \centering
%          \includegraphics[width=\textwidth]{images/lgbm-structure.png}
%          \caption{LightGBM Structure (Vertically Growing)}
%          \label{fig6}
%      \end{subfigure}
%      \hfill
%      \begin{subfigure}[b]{0.20\textwidth}
%          \centering
%          \includegraphics[width=\textwidth]{images/other-methods.png}
%          \caption{Other Boosting Methods Structure (Horizontally Growing)}
%          \label{fig7}
%      \end{subfigure}
% \caption{Boosting Methods}     
% \end{figure}

We use this algorithm on our dataset of 10,377 single-cell images. We split the dataset into training and testing periods, with 80\% training set and 20\% testing steps. For training optimization, we used two types of boosting approaches: ‘gbdt’ commonly used as a default option that uses the decision tree approach, and ‘rf’ is used to increase accuracy based on the random-forest method. We tuned the default parameters according to our dataset. The number of classes was 3, maximum depth of 7 was used with the number of leaves equal to 13. The number of estimators was 50, with a total of 100 iterations. We first trained the model with the default boosting type ‘gbdt’ at the learning rate of 1e-2, which performed well with the training accuracy of 72\% but also caused over-fitness. To improve the results, we then used random-forest as ‘rf’ as our boosting approach. To minimize over-fitness, we make use of bagging by setting a bagging fraction and bagging frequency with L2 regularization. We also did the feature sub-sampling by setting the feature fraction as 0.05, meaning LightGBM would randomly select 5\% of parameters in each iteration to build the tree. To improve accuracy, we used a binning technique with the value of 255 as a maximum bin. The model trained faster than we expected in 4.4 seconds even after using a higher binning value. In addition, we found that the model was not causing over-fitness anymore, and the test accuracy also increased to 75\%.

\subsubsection{Pre-Trained Deep Neural Network Models}
In this section, we report the detailed analysis of the pre-trained models. Different types of deep CNN architectures were adopted such as InceptionResNetV2, ResNet variants, DenseNet and EfficientNet variants  \cite{kassania2021automatic, tan2019efficientnet}. These models are developed particularly for image classification problems. However, each of them differs in terms of how they take inputs and optimize their parameters. We found that during the training period, most of these models performed poorly. Also, these pre-trained CNNs are built on a much bigger dataset while our dataset is relatively smaller for computation. Consequently, the experimental results in most of them give less accuracy, yield high training loss, and cause over-fitness. Among these CNNs, ResNet50V2 was the best model with 69\% accuracy and no ‘overfitting’. To improve the retained result, we fine-tuned the ResNet50V2 according to our dataset. 
We used the original pre-trained ResNet50V2 deep CNN-based model with a custom head but leaving off the fully-connected layers. For the custom layer head, three layers were added on top of the ResNet50V2 model, a global average pooling layer followed by a drop-out and a dense layer. Following the drop-out  layer, a softmax activation function was then applied for the final learning to obtain the prediction.
For training optimization, Adam optimizer was used at the learning rate of 1e-2, which then trained for 15 epochs with the batch size of 16, used for all the pre-trained models. As a loss function, categorical cross-entropy is used, it’s computed by the Eq. \ref{eq1} as follows:
\begin{equation}\label{eq1}
   Loss = - \sum_{i=1}^{\text{output size}} x_{i} *\log \hat{x}_{i}
\end{equation}
% \[Loss = - \sum_{i=1}^{output size} xi *\log \hat{x}i\] \newline

Where $\hat{x}_i$ is the $i^{th}$ scalar value, $x_i$  is the corresponding target value, and the output size is the number of scalar values in the model output.
However, to reduce the chances of overfitting and training loss, we tested the models at different learning rates and selected the one where it performed best. As a result, the model resulted in improvement and was able to achieve 72\% test accuracy at the learning rate of 1e-2.

\subsubsection{Proposed Custom Deep CNN Architecture}
\begin{figure}
\centering
  \includegraphics[scale=0.4]{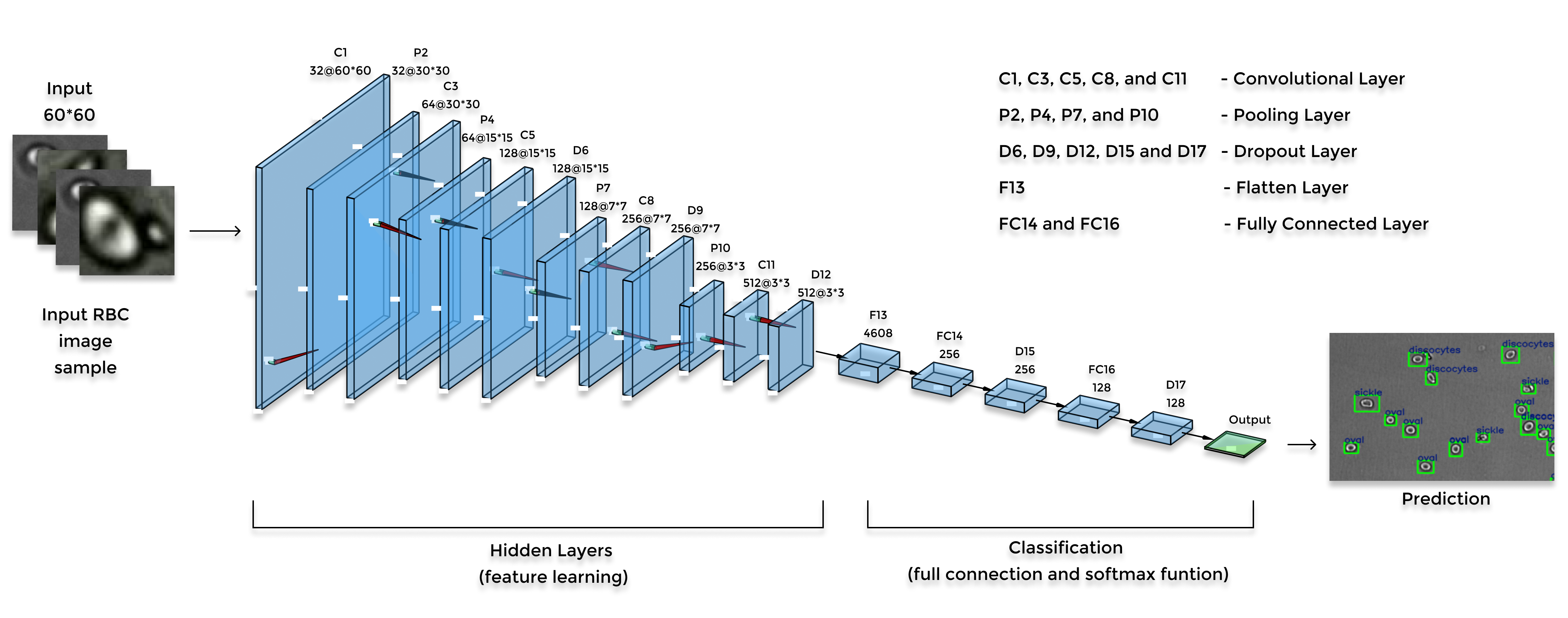}
\caption{Custom Deep CNN Architecture}
\label{fig8}
\end{figure}

After computing and analyzing the Pre-trained CNN models along with the Light GBM algorithm, we propose a deep CNN architecture with 18 layers including 5 convolutional layers (C1, C3, C5, C8, and C11), 4 pooling layers (P2, P4, P7, and P10), 5 dropout layers (D6, D9, D12, D15 and D17, with p = 0.3), flatten layer (F13) and 2 fully connected layers (FC14 and FC16). For the activation function, the ReLU non-linear function is used. Following the D17 layer, a softmax function with a categorical cross-entropy loss function is then applied for the final learning to obtain the final training and predictions. More details about this custom deep CNN architecture are shown in Fig. \ref{fig8}.

Next, we fed our dataset of 10, 377 image samples into the custom deep CNN. We found that the training on this model is more stable than any of those other pre-trained CNNs. To further demonstrate our model's performance, we use some performance metrics like accuracy, precision, recall, and f1-score. Evaluating these metrics is essential as they give us the ground truth about the model. All the details on performance metrics can be seen in Table.\ref{table1} along with the confusion matrix.  

\section{Experimental Results}
\label{sec5}

\subsection{Performance Metrics}
\begin{figure}
\centering
  \includegraphics[scale=0.6]{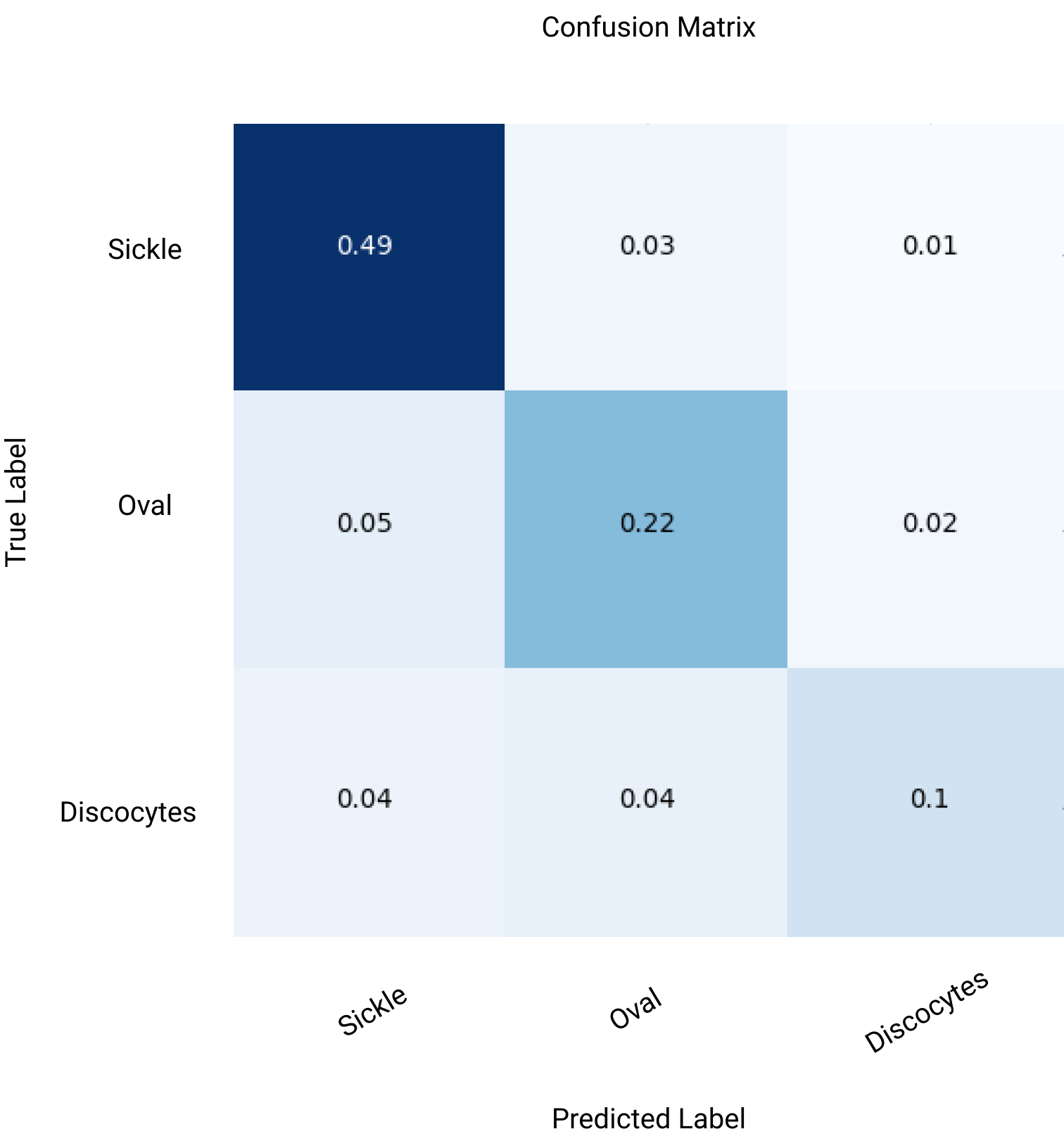}
\caption{Confusion Matrix of Proposed Custom Deep-CNN}
\label{fig9}       
\end{figure}

To evaluate the performance of our proposed model, we use several metrics such as accuracy, precision, recall, f1 score, and confusion matrix.  
% Accuracy - The accuracy here refers to the accuracy of the testing set performed on our proposed model. Precision measures the exactness of classifiers. Having a low precision score can mean a high number of false positives. Recall measures the true positive rate and overall sensitivity in classifiers. Having low recall means the classifier has a large number of false negatives. F1-score analyzes the scores of both precision and recall and gives the best accuracy achieved. Having an f1-score closer to 1 means great accuracy and having a score closer to 0 means the worst accuracy achieved.

\begin{equation}
Accuracy = \dfrac{TP+TN}{TP+TN+FP+FN}
\end{equation}

\begin{equation}
Precision = \dfrac{TP}{TP+FP}
\end{equation}

\begin{equation}
Recall = \dfrac{TP}{TP+FN}
\end{equation}

\begin{equation}
F1 score = \dfrac{2*(Precision*Recall)}{(Precision+Recall)}
\end{equation}

\begin{center}
  \begin{tabular}{lclcl}
	\hline
    &	Positive	& Negative	\\
	\hline
    Positive	& TP(True Positive)	& FP(False Positive)	\\
	\hline
    Negative	& FN(False Negative)	& TN(True Negative)	\\
	\hline
  \end{tabular}
\end{center}

We used all the mentioned metrics to evaluate the performance of the models. Results obtained from the different models can be seen in Table. \ref{table1}.

\begin{table}[ht]
%\resizebox{\columnwidth}{!}{%
\begin{adjustbox}{width=\textwidth}
\begin{tabular}{|l|c|c|c|c|c|c|c|c|c|c|}
\hline
\multirow{3}{*}{\textbf{\diagbox{Model}{Metric}}}                     & \multicolumn{10}{c|}{\textbf{Performance Metrics}}                  \\ \cline{2-11} 
& \multirow{1}{*}{Accuracy} & \multicolumn{3}{l|}{F1-Score} & \multicolumn{3}{l|}{Precision} & \multicolumn{3}{l|}{Recall} \\ \cline{3-11} 
&          & Sickle  & Oval  & Discocytes  & Sickle   & Oval  & Discocytes  & Sickle  & Oval & Discocytes \\ \hline
InceptionResNetV2                                        & 0.67                      & 0.81    & 0.53  & 0.08        & 0.69     & 0.58  & 0.90        & 0.97    & 0.49 & 0.04       \\ \hline
VGG16                                                    & 0.54                      & 0.70    & 0.04  & 0.00        & 0.54     & 0.47  & 0.00        & 1.00    & 0.02 & 0.00       \\ \hline
VGG19                                                    & 0.52                      & 0.69    & 0.00  & 0.00        & 0.53     & 0.00  & 0.00        & 1.00    & 0.00 & 0.00       \\ \hline
ResNet50                                                 & 0.69                      & 0.84    & 0.60  & 0.10        & 0.80     & 0.52  & 0.83        & 0.88    & 0.72 & 0.05       \\ \hline
ResNet50V2                                               & 0.70                      & 0.84.   & 0.60  & 0.26        & 0.77     & 0.57  & 0.73        & 0.93    & 0.63 & 0.16       \\ \hline
ResNet50V2$^\dag$                                    & 0.72                      & 0.84    & 0.59  & 0.45        & 0.77     & 0.61  & 0.65        & 0.92    & 0.54 & 0.32       \\ \hline
DenseNet 121                                             & 0.61                      & 0.76    & 0.42  & 0.00        & 0.64     & 0.50  & 0.00        & 0.95    & 0.36 & 0.00       \\ \hline
EfficientNet B0                                          & 0.53                      & 0.69    & 0.00  & 0.00        & 0.53     & 0.00  & 0.00        & 1.00    & 0.00 & 0.00       \\ \hline
LightGBM                                                 & 0.75                      & 0.86    & 0.63  & 0.60        & 0.82     & 0.71  & 0.57        & 0.89    & 0.56 & 0.63       \\ \hline
Custom Deep CNN                                          & 0.81                      & 0.89    & 0.75  & 0.65        & 0.84     & 0.75  & 0.77        & 0.94    & 0.73 & 0.55       \\ \hline
\end{tabular}%}
\end{adjustbox}
\caption{Performance Comparison of our customised Deep CNN along with baseline models on test samples, where $^\dag$ depicts the a fine-tuned version of ResNet50V2}
\label{table1}
\end{table}

% \section{Testing Experiments}
\subsection{Cell Detection}
\begin{figure*}
\centering
  \includegraphics[scale=0.27]{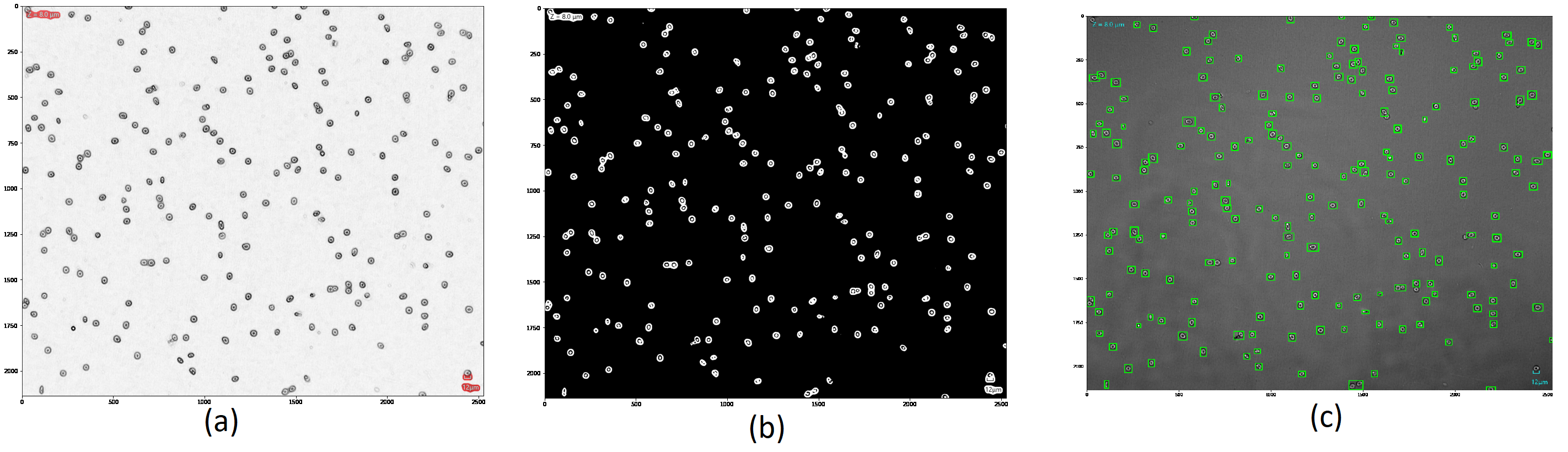}
\caption{(a). Dilated Image, (b). Dilated image converted into a binary image, (c). Detection of Separated Components}
\label{fig10}

\end{figure*}

Image morphological transformation methods are used to detect all the cells in one particular blood slide image sample. There are two basic morphological transformations, Erosion and Dilation, which can be used to remove noises, finding individual elements, intensity bumps or holes in an image. As Erosion method tends to remove the noises and the boundary of the elements in the image it was removing the cell boundaries as well, so only Dilation operator was used to find distinct cell in the image. The given image was then dilated and converted into a binary image as shown in Fig. \ref{fig10} (a, b). Afterwards, using the binary image, all separated components were cropped out as single cells. To separate the clusters of cells, the process was repeated, taking the clustered area as the initial image. Eventually, all the cells were detected in the given blood slide image, as shown in Fig. \ref{fig10} (c).

As explained in the previous section, the proposed custom deep CNN model performed better than the rest of the models in the training period. Now, we test it using a random image as a test image sample. We use this image (Fig. \ref{fig10} c) to test and analyze the performance of our proposed model. An experiment was conducted. For the inspection, we tested solely the three classes of sickle anemia disease that we used - sickle, oval, and discocytes.

\subsection{Cell Prediction}
\begin{figure}
\centering
  \includegraphics[scale=0.2]{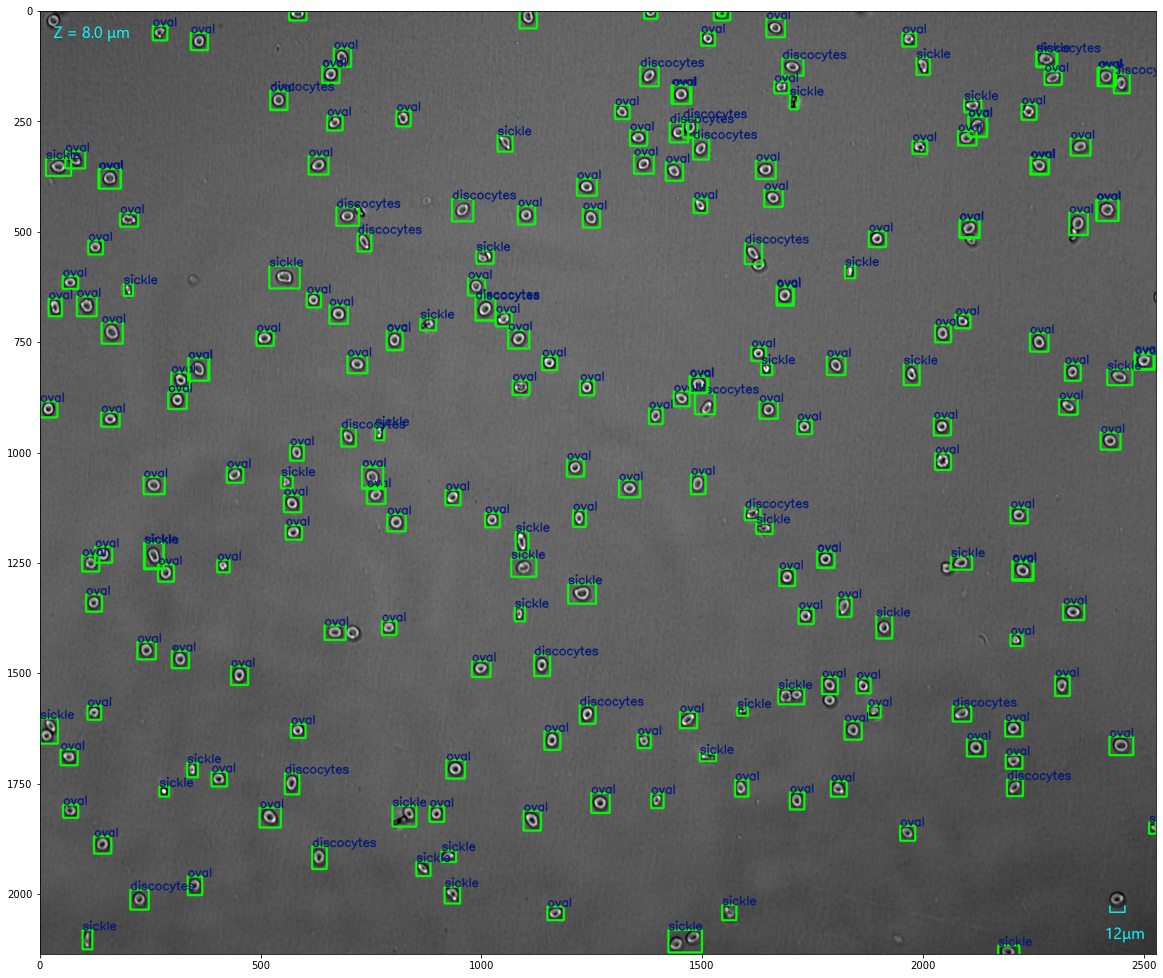}
\caption{Cell Prediction from the proposed custom deep CNN architecture.}
\label{fig11}

\end{figure}

For prediction, we fed a test image that consists of multiple cells in a single image (from the original 428 images) to our trained deep CNN model. The image used for this test can be seen in Fig. \ref{fig10} c. We found that the cells from all three classes (Sickle, Oval, and Discocytes) were detected in the test image. The prediction in the test image can be seen in Fig.\ref{fig11}.

\subsection{Results}
In this section,  experimental results are reported of the proposed Custom Deep CNN Model and other models.  We considered 428 raw microscopy annotated images. Based on these annotated images, 10,377 single RBC image samples were extracted using the cell segmentation and cell detection approach explained in section \ref{sec4}. Subsequently, these 10, 377 single-cell images were normalized to the same size (60*60). These new single-cell image datasets were assigned to 3 different categories namely; sickle, discocytes, and oval. A comparison study on all the models (Light GBM and 9 deep CNN-based models) was conducted to observe and obtain the best results. We used 9 CNN-based models namely, InceptionResNetV2 \cite{kassania2021automatic}, VGG16 \cite{simonyan2014very}, VGG19 \cite{simonyan2014very}, ResNet50 \cite{kassania2021automatic}, ResNet50V2 \cite{kassania2021automatic}, Fine-tuned ResNet50V2 \cite{kassania2021automatic}, DenseNet \cite{kassania2021automatic}, EfficientNetB0 \cite{tan2019efficientnet}, and a custom deep CNN-model. An analysis using LightGBM \cite{ke2017lightgbm} was also conducted. The LightGBM algorithm has been known famously for its performance compared to state-of-the-art algorithms. After computing and evaluating all of the models, the custom deep-CNN model performed much better than the rest with 81\% test accuracy. Apart from the `Custom Deep CNN' model, the `Fine-tuned ResNet50V2’ and 'Tuned LightGBM' models had slightly better results than the remaining models with 72\% and 75\% test accuracy. Detailed figures of the performance matrix along with the confusion matrix are shown in Fig.\ref{table1} and Fig. \ref{fig9}.

\subsection{Interpretability}
In this section, we investigate our proposed model’s interpretability using two different algorithms called LIME (Locally Interpretable Model-agnostic Explanations) \cite{lime} and SHAP (SHapley Additive exPlanation) \cite{NIPS2017_7062}. Both of these algorithms create a surrogate model by changing the input slightly and testing the changes in the prediction. We use LIME because it tries to solve the model interpretability of any classifier by producing faithful explanations. However, LIME provides transparency and accountability. It lacks some consistency and missingness, meaning it surrogate model locally and around an individual input. Whereas, SHAP fulfills these properties using the classic ‘Shapley Values’ from the game theory concept to explain the output of a black-box and provide the theoretical guarantees. 

For the deep neural networks, one has to make strenuous efforts to interpret the behavior of the predictions using the black-box model, especially in terms of medical imaging. Therefore, model interpretability is of utmost importance.

\subsubsection{LIME}
\begin{figure*}
\centering
  \includegraphics[scale=0.4]{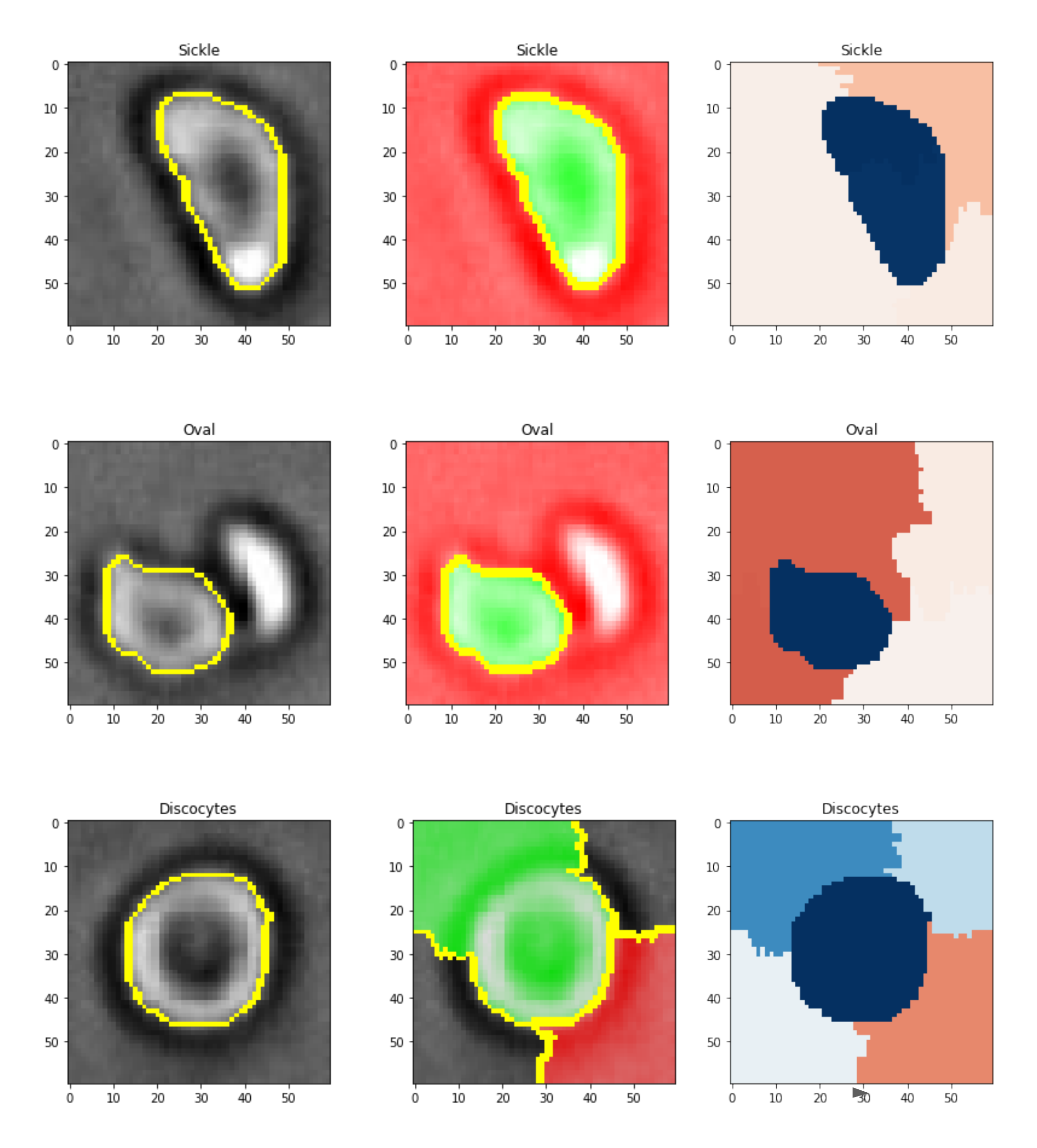}
\caption{LIME Saliency Maps for Sickle, Oval and Discocytes.}
\label{fig12}       
\end{figure*}

We first employ LIME for the model interpretability of the predictions made by our proposed model. The single-cell images of 'Sickle', 'Oval', and 'Discocytes' cells were fed as the input image to the LIME explanation in order to generate perturbations by superpixels over a number of samples (1000 samples were taken to determine the artificial data points that are generated by LIME). As it interprets the model at a local region in the feature space,  saliency maps extracted from the LIME can be seen in Fig. \ref{fig12}. It shows the regions of the distorted cells having the greatest influence on the prediction result using our proposed model. To begin with, the mask boundaries generated for all three types of cells can be seen in the yellow region. In addition, we can see how our model classifies an image as a particular cell by segmenting the predicted samples into superpixels. Furthermore, after the yellow region image, we can see the 'pros and cons' image of the same, colored in green and red region. Both green and red region defines the detected superpixels that contributed to the prediction. Lastly, the third image is visualized as a heatmap where the detected superpixels are in blue and the rest of the insignificant region in light-to-dark color.

\subsubsection{SHAP}
SHAP has an additive attribution property, and it computes the explanations using shapely values - every single feature is assigned for a particular prediction. We use a similar approach as we used earlier in LIME. The single-cell images of `Sickle', `Oval', and `Discocytes' cells were fed as the input to generate perturbations. To build a partition explainer with the proposed model, an image masker was defined to mask out the partitions of the input image for blurring and inpainting. A total of 500 evaluations were used to estimate the SHAP values and to get the explanations with sufficient granularity for the superpixels. Interpretations of SHAP explanations through blurring and inpainting can be seen in Fig. \ref{fig12} and Fig. \ref{fig13}, showing how the model classifies an image as a particular cell using the SHAP values and superpixels. As we can see in Fig. \ref{fig12}, the three rows comprise the SHAP explanations for the sickle, discocyte, and oval cells through blurring. The first row shows the original input image of the sickle cell followed by the three probable predictions. For the sickle cell, the first prediction is classified as a sickle, followed by the next probable classes being discocytes and oval. The reason behind its classification as a sickle cell is because of the SHAP values of the superpixels, which can be seen highlighted in a red region. Whereas the blue region seen is the next probable class to base predictions. The next two rows show the probable predictions in the same manner. Similarly, Fig. \ref{fig13} shows the interpretations of SHAP explanations through inpainting with all three images— classified as sickle, oval, and discocytes along with their corresponding three probable predictions classes. After using the blur values for the masked region (or patch), we filled the masked parts through inpainting. The masker used is the image masker with inpaint-telea, which interprets the SHAP explanation more visibly clear.

\begin{figure*}
\centering
  \includegraphics[scale=0.4]{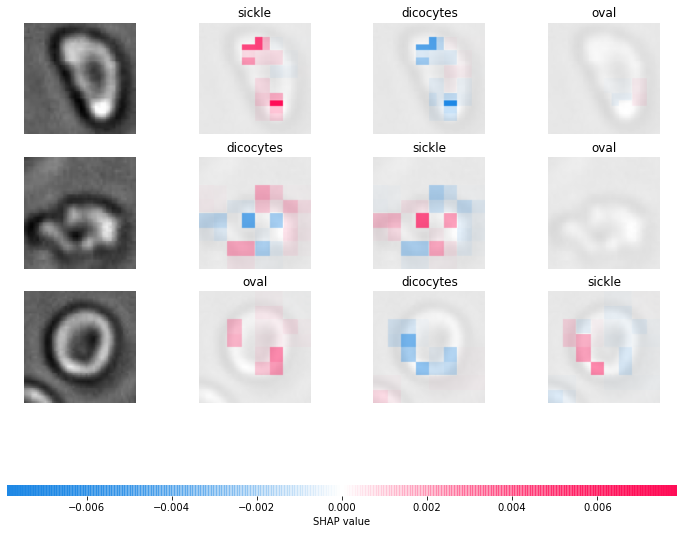}
\caption{Interpretations of SHAP Explanations through Blurring.}
\label{fig13}       
\end{figure*}

\begin{figure*}
\centering
  \includegraphics[scale=0.4]{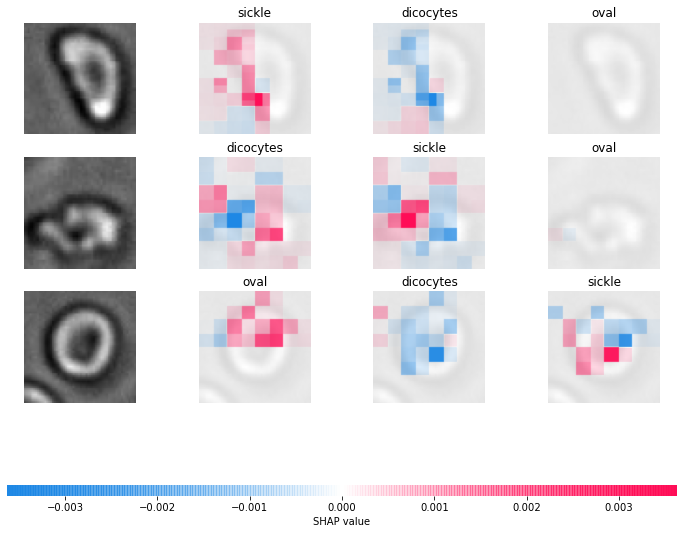}
\caption{Interpretations of SHAP Explanations through Inpainting.}
\label{fig13}       
\end{figure*}

\section{Case Study}
There are a limited number of ways we can adopt to diagnose Sickle Cell Disease (SCD). If we categories them we will end up with two broad sets of categories. One category can be the traditional methods and technologies being used for diagnosing SCD, and the other category can be classified as advanced modern techniques emerged from the advancement and application of artificial intelligence in medical practices. These categories are discussed as follows.

\subsection{Traditional methods for detecting SCD}
\subsubsection{High-performance liquid chromatography (HPLC):} A technique used to separate the haemoglobin fraction, detect and quantify Hb variants has been developed \cite{arishi2021techniques}. Different variants of Haemoglobin are separated according to their retention value on a stationary phase for the diagnosis of SCD. Some Haemoglobin variants have the same retention value which makes it difficult to identify the exact variant which can lead to wrong diagnosis. Hence, this method is good to be used as a supplementary test.    
\subsubsection{Genetic Test:} A medical test done to identify the changes in genes. In some cases, the patients are tested based on a genetic study for the precise detection of different classes of SCD erythrocytes based on $\beta$ - globin mutation \cite{arishi2021techniques}. A gene theory is done widely for hemoglobinopathy, otherwise known as to detect the hemoglobin abnormalities. A patient is diagnosed with SCD based on the detection of the number of Hb S gene mutations [3]. 
\subsubsection{Supportive Care:} In critical situations, the patients are given blood transfusion to lower the Hb levels according to their needs \cite{vichaka2015differentiation}. Other intensive methods are, to do the complete sickle blood cell count, or bone marrow transplant \cite{vichaka2015differentiation}. As of now, the only way to avoid transfusions for the patients is to get the allogeneic hematopoietic stem cell transplantation therapy \cite{boraldi2020rare}.

\subsection{Advanced methods for detecting SCD} 
\subsubsection{Segmentation Techniques:} AI Tools for cell segmentation by RSIP Vision \cite{ben2017retinal} have been developed to classify abnormality in RBCs. Their method recognizes the cell based on their morphology. In addition, their software provides methods for building an awareness system. In our work, the developed architecture can easily quantify the distorted SCD erythrocytes into different classes recognizing  not only the abnormality but also defining the class group a distorted cell belongs to. 
\subsubsection{Smartphone-based technique:} A rapid screening test using a smartphone microchip, a microscope and machine algorithm was developed by De Haan et al. \cite{de2020automated}. Module consisted of two NNs, one to enhance the picture taken and second to perform semantic segmentation. The method developed was affordable and easy to use. However, the method is good for binary results as it cannot classify sickled RBCs in different classes. 
\newline

The case study stated above reveals that regardless of traditional medical practices, some AI tools have been developed and are available to use for erythrocytes classification. Furthermore, this discussion describes the necessity of developing a method for multi-class classification. 
\newline

Specifically, scientists are progressively creating new mediums to detect SCD in depth for better assessment. In many places, the treatments for sickle cell disease are yet to be established. Therefore, an assessment using segmentation approaches can be done before going for medical care and therapy. Hence, in this context, our proposed method can be used by the clinicians even in the undeveloped and poor environment areas.

\section{Discussion}\label{discussion}

The current study is a small step towards improving identification criteria through quantification and identification of Erythrocytes, i.e., Red Blood Cells, using deep CNN methodology, which will help identify and diagnose Sickle Cell Disease early to contribute towards existing technology. This research has value not just for the single disease, if we could identify and quantify cells related to single pathology, the same can be replicated for other morphological pathology changes related to other diseases like Malaria, Iron Deficiency Anaemia or dengue or any other infectious disease or medical condition that involves morphological changes and their morbid affects on human physiology.\\
Cell counters and analyzers can be modified or upgraded through this research not to just count and identify normal blood cells but to identify and indicate flagging or alarming the interpretation of reports for screening activity of any medical condition based on particular shape changes. The changes can be identified if we have database of shapes and morphological modifications, which can be compared at one point. This can be helpful in identifying the linked morphology to a particular clinical condition. 

This research has scope much wider than what we are able to find through our study as more data points we can gather, more cells images we can annotate, better will be the utility of the overall information. This will require resources, more scientifically trained human resources and more time to explore the segmentation criteria so that all aspects of the disease condition can be considered and studied in detail. However, this study is just a starting point for the larger aim of making medical diagnosis and screening as close to real-time medical examination as possible, to save the time of our medical experts who can spend more time with their patients rather than looking under the microscope just to find the correct cells with relevant morphology.

In our experiments, we did some robust object classification using only 428 images taken from SCD patients. 
Versatile and reliable classification from medical images is a pivotal step for biomedical research. A major challenge for image classification in medical imaging is the image quality, distorted cells, variations of distorted cells with the normal cells, along with the artifacts in the whole image. While addressing this issue, our method consists of multiple steps involving data pre-processing, image segmentation, and deep learning classifier models. Our model can be used in multiple classification problems as a combination of different classification and detection modules.

\section{Conclusion}
In this work, a deep convolutional neural network-based model is proposed to identify and quantify the distorted and normal morphology of erythrocytes. We generated a dataset using the 428 raw microscopic images taken from the blood samples of Sickle Cell Disease (SCD) patients. We focused on the three well-defined erythrocyte shapes that are discocytes, oval, and sickle. We performed the multi-class classification using the baseline models composed of neural network-based architectures. A total of 9 different deep neural network-based models were employed. For model comparison, a tuned Light Gradient Boosting Machine (LightGBM) was also implemented. Our proposed custom deep-CNN model performed better with 81\% accuracy and better precision, recall, and f-score than the rest of the baseline models. We also investigated our proposed model’s interpretability using LIME and SHAP. 

In the future, we intend to experiment on a much larger dataset to check the capability by integrating vision transformer and other attention network. We can improve our dataset by scaling it to more rare categories, e.g., granular or stomatocytes, etc. Improving the dataset can also help us improve the model to identify more erythrocyte shapes. Results in our study signify that \ac{ML}  can be useful for clinicians for quick and accurate analysis of SCD blood samples.

\section*{Acknowledgment}
This work was supported by the Academy of Finland (grants 336033, 315896), Business Finland (grant 884/31/2018), and EU H2020 (grant 101016775).

\bibliography{references}

% \appendix

\end{document}